# Magnetic Droplet Soliton Nucleation in Oblique Fields

Morteza Mohseni,[1, †, §] M. Hamdi,[1, †] H. F. Yazdi,[1] S. A. H. Banuazizi,[2] S. Chung,[2,3] S. R. Sani,[2,3] Johan Åkerman,[2,3] and Majid Mohseni[1, *]

[1] *Faculty of Physics, Shahid Beheshti University, Evin, Tehran 19839, Iran*
[2]*Materials and Nano Physics, Department of Applied Physics, School of Engineering Sciences, KTH Royal Institute of Technology, Electrum 229, SE-164 40 Kista, Sweden*
[3] *Department of Physics, University of Gothenburg, Fysikgränd 3, 412 96 Gothenburg, Sweden*

We study the auto-oscillating magnetodynamics in orthogonal spin torque nano-oscillators (STNOs) as a function of the out-of-plane (OOP) magnetic field angle. In perpendicular fields and at OOP field angles down to approximately 50 degrees we observe the nucleation of a droplet. However, for field angles below 50 degrees, experiments indicate that the droplet gives way to propagating spin waves, in agreement with our micromagnetic simulations. Theoretical calculations show that the physical mechanism behind these observations is the sign changing of spin-wave nonlinearity (SWN) by angle. In addition, we show that the presence of a strong perpendicular magnetic anisotropy (PMA) free layer in the system *reverses* the angular dependence of the SWN and dynamics in STNOs with respect to the known behavior determined for the in-plane magnetic anisotropy free layer. Our results are of fundamental interest in understanding the rich dynamics of nanoscale solitons and spin-wave dynamics in STNOs.

## I. INTRODUCTION

Solitons are localized waves that preserve their shape during their time evolution and can form when nonlinear effects overcome wave dispersion. They can occur in a range of physical systems, including photonic [1–5], phononic [6–9], and magnonic [10–14] systems. Magnonics [15–18], which is concerned with studying spin waves, has received significant research interest in recent decades due to its ability to serve as an alternative technology in future electronic devices in information processing units [19–22]. After the pioneering work of Slonczewski [23–25], the spin transfer torque effect (STT) in spin torque nano-oscillators (STNOs) is known to be a very powerful technique for generating and controlling exchange-dominated spin waves (SWs) at microwave frequencies by injecting a DC current through a nanocontact (NC) in pseudo-spin-valve (PSV) structures [26].

The generation and detection of different types of SWs in STNOs—such as propagating SWs [23–25,27], localized SW bullets [10–12,28,29], vortices [30–33], dynamical skyrmions [34], and magnetic droplet solitons (henceforth "droplets") [14,35–44]—has been demonstrated and studied extensively from both experimental and theoretical points of view. Droplets were introduced in 2010 [13] as the dissipative counterpart of conservative magnon drops [45,46], and were observed experimentally for the first time in 2013 [14,35]. Although the various rich dynamical aspects of these nanoscale solitons—such as drift resonance instability [42,47], propagation of droplets [44], generation with spin currents [48], nucleation boundaries and its temperature dependence [40,49], merging [41], and perimeter mode excitation [43] and direct observation of droplet by scanning transmission x-ray microscopy [50]— have been studied since their initial observation, all the research has confirmed the bare possibility of stable droplet nucleation in oblique fields [13,47].

It is well known from the theory of droplets that, when the applied field is canted—even at small angles—nucleated droplets tend to escape from the NC due to the drift instability process, since the in-plane component of the effective field dislodges the droplet from its initial position [13]. As a consequence, it is difficult for droplets to remain stable in oblique fields, although the evidence we present here challenges this assumption, or at least demonstrates that the drift instability does not necessarily has to take place even when the relevant conditions are fulfilled.

Here, we study the angular dependence of the magnetization dynamics in orthogonal STNOs using experiments, micromagnetic simulations, and theoretical calculations. We show that the nucleation of droplets is not limited to perpendicular fields, and that droplet nucleation in oblique fields at a critical angle can be observed both through the magnetoresistance (MR) and the magnetodynamics of the STNO. Using the fact that the frequency of a droplet mode always lies below the ferromagnetic resonance (FMR) frequency [13], we determined the critical angle $\theta_c$ at which droplet solitons convert to propagating SWs. Theoretical calculations of the angular dependence of the nonlinear frequency shift parameter confirm our experimental evidence by indicating that the sign of the nonlinearity changes at certain angles, which means that the nucleation of a soliton is possible for the system. Interestingly, this also demonstrates that the presence of a strong perpendicular magnetic anisotropy (PMA), *reverses* the angular dependence of the SW nonlinearity in STNOs.

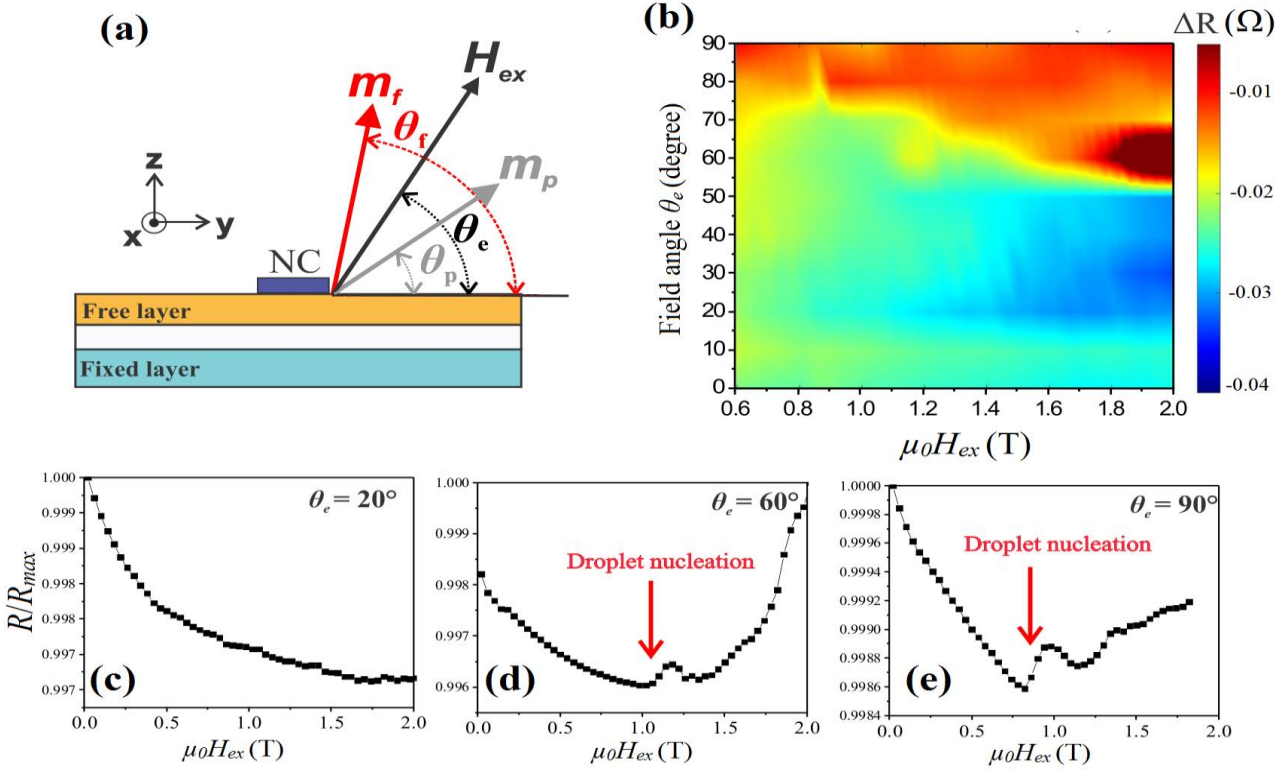

Figure 1. (a) Schematic of device structure and geometry with the external field angle $\theta_e$, angle of the fixed layer magnetization $\theta_p$, and angle of the free layer magnetization $\theta_f$, (b) color plot of the measured angular dependent $\Delta R = R(\mu_0 H_{ex}) - R(\mu_0 H_{ex} = 0\,T)$ of the system at a constant current of $I_{dc} = 20$ mA; field-dependent normalized (to its maximum) MR at $I_{dc} = 20$ mA and (c) $\theta_e = 20°$, (d) $\theta_e = 60°$, (e) $\theta_e = 90°$.

## II. METHODS

The samples investigated in this study are orthogonal pseudo spin valves (PSVs) of Co8/Cu8/Co0.3[Ni(0.8)/Co(0.4)]$_{\times 4}$ (numbers are thicknesses in nm) used in [14,40] with a NC radius of 50 nm, schematically shown in Figure 1(a). Details of the sample preparation and our experimental methods can be found in Ref. [14,40]. All measurements were performed at room temperature in the presence of an accurate and uniform rotatable magnet. The DC current applied to the NCs had a negative polarity, with electrons flowing from the free to the fixed layer.

All micromagnetic simulations were performed using GPU-based MuMax 3.0 core [51] in a cuboid-like volume for the free layer with a size of 1000 nm × 1000 nm × 3.6 nm, divided into 256 × 256 × 1 cells. Under these conditions, cell sizes are well below the exchange length of the free layer [14]. An infinite cylinder-like NC with a radius of 50 nm was defined at the center of the device for spin-polarized current driven excitations, which is supplemental to the effects of the Oersted field. In all the simulations and theoretical work, we used $\mu_0 M_{s,f} = 0.9$ T, $\alpha = 0.03$, $\mu_0 H_k = 1.2$ T and $A_{ex} = 12.3$ pJ/m as the saturation magnetization, Gilbert damping, PMA field and exchange stiffness for the free layer, respectively [14]. Absorbing boundaries were used to prevent back-reflection and interference effects [52]. Spin polarization, which is defined thorough the magnetization angle of the fixed layer ($\mu_0 M_{s,p} = 1.7$ T) $\theta_p$, was found by solving the magnetostatic boundary conditions for the fixed layer by scanning the external field amplitudes at different angles $\theta_e$ (see Appendix B and Ref. [53]).

## III. RESULTS AND DISCUSSION

### A. Experiments and micromagnetic simulation

The MR versus field at different angles and with a constant applied current of $I_{dc} = 20$ mA was measured in order to detect the nucleation of a droplet. Figure 1(b) represents a color map of the measured MR of the system, which indicates that the MR starts to increase from certain angles when the system tends to transform from small-angle precession to a droplet. Below $\theta_e \sim 50°$, there is no sign of a droplet at any field. The field-dependent MR of the STNO at different angles of $\theta_e = 20°$, $\theta_e = 60°$ and $\theta_e = 90°$ with a constant current of $I_{dc} = 20$ mA is presented in Figure 1(c)–(e). At $\theta_e = 20°$, the MR decreases with increasing field, behaving as expected in orthogonal PSVs [54]. For $\theta_e = 60°$ to $\theta_e = 90°$, MR first decreases, but at a certain field this trend changes and the MR increases, indicating the formation of a droplet under the NC [14].

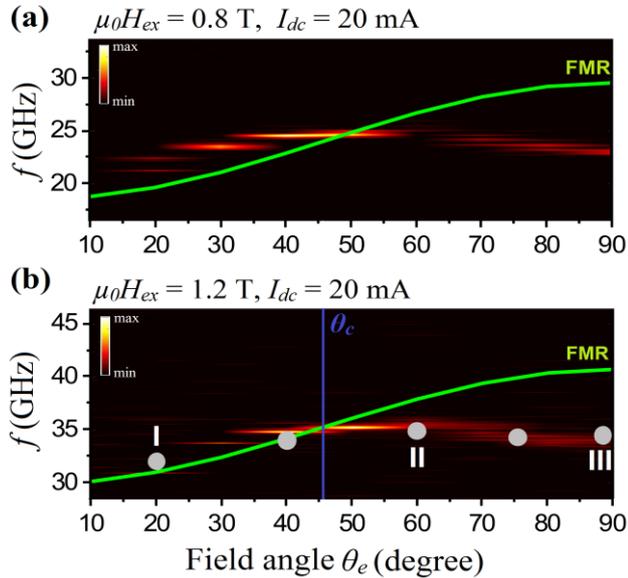

Figure 2. STNO frequency measured vs. applied field angle for the applied current of $I_{dc} = 20$ mA: (a) $\mu_0 H_{ex} = 0.8$ T and, (b) $\mu_0 H_{ex} = 1.2$ T; solid green curves show the angular-dependent FMR frequency while the solid blue line indicates the critical angle for droplet nucleation. Solid gray circles in (b) show the results from micromagnetic simulations and I, II, III indicate respectively the selected angles of $\theta_e = 20°$, $\theta_e = 60°$, and $\theta_e = 90°$ for analyzing the spatial profile of the dynamics, presented in Figure 3.

The angular dependence of the STNO frequency measured at a constant current of $I_{dc} = 20$ mA is shown in Figure 2. The top row (a) shows the results for an applied field of $\mu_0 H_{ex} = 0.8$ T while the bottom row (b) shows the results for an applied field of $\mu_0 H_{ex} = 1.2$ T. The solid green lines show the angular-dependent FMR frequency of the system (See Eq. B18 of Appendix B). As droplet frequency always lies below the FMR [13] and considering the MR measurements, the critical angle of droplet nucleation is defined as the angle where the frequency of the system drops below the FMR frequency, which is indicated by solid blue vertical line in Figure 2(b). Solid gray circles in Figure 2(b) show the results from micromagnetic simulations which exhibit a good agreement with experiments. Labels I, II, III in Figure 2(b) mark points in the spectrum of the STNO for which we present the spatial profiles of the excited wave obtained by micromagnetic simulations at angles of 20°, 60°, and 90° in Figure 3(a)–(c), respectively.

Two different modes exist in the system. Up to a definite critical angle, the STNO frequency is *higher* than the FMR frequency, reflecting a propagating nature of the generated SWs [55]. Above this angle, where the STNO frequency drops below the FMR frequency, excited spin waves demonstrate a localized non-propagating solitonic nature [13]. This critical angle is indicated by vertical blue line in Figure 2(b). It should be noted that, up to $\theta_e = 10°$, we could not detect any experimental signal, since the power of the STNO is below the detection resolution of the instruments, probably due to the small precession angle [29].

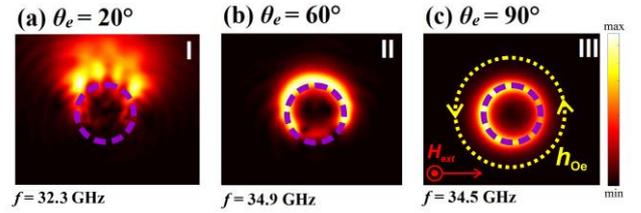

Figure 3. Spatial profile of the SW dynamics at the applied current of $I_{dc} = 20$ mA and external applied field of $\mu_0 H_{ex} = 1.2$ T; (a), (b), and (c) show the spatial profiles of the modes at $\theta_e = 20°$, $\theta_e = 60°$, and $\theta_e = 90°$, respectively, as indicated by I, II, and III in Figure 2(b).

In order to better understand the nature of these modes, the spatial distribution of the amplitude (FFT) of the excited spin waves obtained by micromagnetic simulations are shown in Figure 3. Three applied field angles of $\theta_e = 20°$, $\theta_e = 60°$, and $\theta_e = 90°$—indicated as I, II, and III in Figure 2(b)—were selected to provide better insight and to evaluate the role of the Oersted field of the DC current on the modes. At the low angle ($\theta_e = 20°$) shown in Figure 3(a), the spatial profile of the spin-wave mode shows a spin-wave beam, which is the result of broken spatial symmetry around the NC induced by the Oersted field of the DC current [56]. This spin-wave beam propagates from above the NC perimeter, where the Oersted field tends to compete with and reduces the in-plane component of the external field. The lower effective field in that region shifts the dispersion to lower frequencies and consequently confines the spatial distribution of the propagating waves [55]. This may be of both fundamental and technological importance in synchronizing STNOs, since spin-wave beams can be employed as the main driving source in this technique [57,58].

By increasing the applied field angle above $\theta_c$, the lower frequency mode disappears and a higher frequency mode begins to appear (as it can be seen from Fig. 2(a-b)), suggesting that a new mechanism begins to dominate the system. Figure 3(b) shows the spatial distribution of the mode at $\theta_e = 60°$. As expected from the electrical measurements (both MR and STNO frequency), the droplet is nucleated at this angle. However, due to the asymmetric energy landscape around the NC induced by the Oersted field of the applied current, the droplet tends to drift toward regions with a lower in-plane field above the NC [13,42]. This drift is not sufficiently high to allow the droplet to run away from the NC. As the canted angle of the external field increases, the energy landscape becomes more symmetric, and finally the drift instability disappears at $\theta_e = 90°$ (Fig. 3(c)), since the in-plane component of the effective field around and outside the NC becomes spatially homogeneous. This leads to the lack of any external forces on the droplet to dislodge it from its initial position. Consequently, at this angle the spatial profile of droplet becomes the most symmetric. However, due to the interplay between the canted fixed layer and the Oersted field of the applied current, the spatial symmetry of the droplet profile remains broken [53].

Based on experimental results, and by applying a similar analysis for a range of the applied fields from $\mu_0 H_{ex} = 0.6$ T to $\mu_0 H_{ex} = 1.2$ T, we obtained a field-angle phase diagram for the angular dependence of the magnetization dynamics in orthogonal STNOs, in order to distinguish the presence of the soliton and the propagating wave when the system is obliquely magnetized.

As indicated in Figure 4, for each field strength below a certain critical angle $\theta_c$, the spin-wave dynamics of the STNO show a propagating wave nature. However, above $\theta_c$, the excited spin waves convert to localized solitonic droplet modes.

Indeed, our results in orthogonal STNOs show an interesting contradiction with the angular trend of spin-wave dynamics in reported systems with in-plane magnetic anisotropy (IMA) [10–12,28,29]. Our evidence demonstrates the existence of propagating waves at nearly in-plane angles, and a droplet soliton in nearly perpendicular angles—while for IMA systems, the solitonic spin-wave bullet exists for in-plane fields and the propagating waves for perpendicular fields.

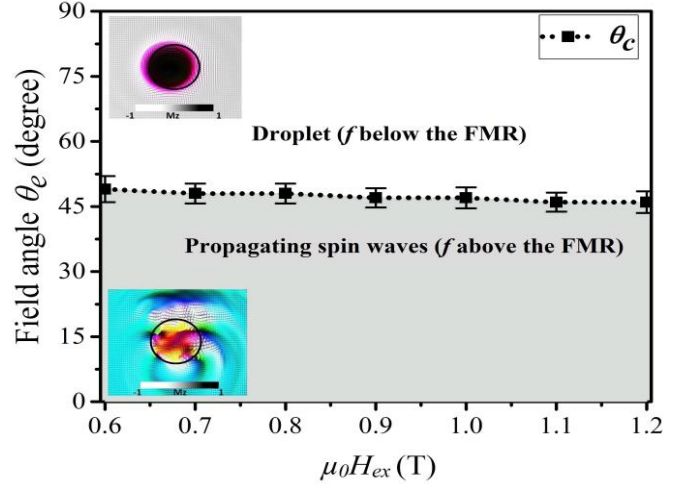

Figure 4. Phase diagram of the droplet nucleation critical angles $\theta_c$ vs. the applied field magnitude and angle, which separates the localized droplet area (white region) and non-localized propagating SW area (gray region).

## B. Theory

In order to describe the physics behind this spin-wave mode conversion, we employed the standard spin-wave Hamiltonian formalism [59,60]. Figure 1(a) shows the STNO-device schematics, where $\theta_e$, $\theta_p$, and $\theta_f$ represent the external field angle and the fixed and the free layer magnetization angles, respectively. We derived an approximate equation for the dimensionless complex spin-wave amplitude $a(\mathbf{r}, t)$ of the variable magnetization which described in Appendix B:

$$i\frac{\partial a}{\partial t} = -\left(\omega_0 - D\nabla^2 + N|a|^2\right)a + i\left(\Gamma_J - \Gamma_0 + \kappa\nabla^2\right)a. \quad (1)$$

Here $\omega_0 = \sqrt{\omega_H[\omega_H - (\omega_K - \omega_M)\cos^2\theta_f]}$ is the linear FMR frequency ($\omega_H = \gamma H_{int}$, $\omega_K = \gamma H_K$, $\omega_M = \gamma M_S$), $\gamma$ is the gyromagnetic ratio, $D = A\omega_M \lambda_{ex}/\omega_0$ is the dispersion coefficient of spin waves due to exchange interaction, $A = \omega_H - \frac{1}{2}(\omega_K - \omega_M)\cos^2\theta_f$, $\nabla^2$ is two dimensional Laplace operator in the film plane, $N$ is the nonlinear frequency shift parameter, $\Gamma_J = [\sigma\cos\gamma_p/(1 + \nu\cos\gamma_p)]\Theta(r - R_{NC})$ where $\gamma_p = \theta_f - \theta_p$ is the angle between the magnetization of the free and the fixed layers, $\Theta(x)$ is the Heaviside step function, $\nu = [\lambda^2 - 1/\lambda^2 + 1]$, in which $\lambda$ is the STT asymmetry factor, $\sigma = I/j_0 R_{NC}^2$, $j_0 = [M_{s,f}^2 e\mu_0 t_f(\lambda^2 + 1)/\hbar\varepsilon\lambda^2]$, $I$ is the applied current through the NC, $\varepsilon$ is the STT efficiency, $t_f$ is the thickness of the free layer, and $\Gamma_0 = \alpha A$ and $\kappa = \alpha\omega_M\lambda_{ex}$. Considering $N$ with respect to the angle of the external field $\theta_e$, there is an essential difference between PMA-STNO and IMA-STNO devices: as shown in Figure 5(inset), unlike in IMA-STNOs, $N$ is positive in PMA-STNOs for in-plane fields; when $\theta_e$ increases, it vanishes at an angle $\theta_{lin}$, changing sign above this angle and becoming negative. This essentially means that, in STNOs, the presence of a strong PMA *reverses* the angular dependence of the spin-wave nonlinearity.

This angular-dependent sign alternation allows the system to satisfy the well-known Lighthill criterion ($ND < 0$) above $\theta_{lin}$, meaning the system is capable of hosting the nonlinear solitonic droplet mode. Thus, as we mentioned, by increasing $\theta_e$ from an in-plane to a perpendicular angle, from $\theta_{lin}$ and above where the nonlinearity changes its sign from positive to negative, it acts against the dispersion. By increasing the applied field angle $\theta_e$ from $\theta_{lin}$ to the critical angle of droplet nucleation $\theta_c$, the nonlinear effects grow; finally $N$ is sufficiently strong at $\theta_c$ to allow the propagating waves to become modulationally unstable and convert to a solitonic droplet mode [13]. Figure 5 shows $\theta_{lin}$ versus field amplitude; this was found by applying the same analysis for different amplitudes of the applied field. By comparing Figure 4 and Figure 5, it can be seen that $\theta_{lin}$ turns out to be smaller than $\theta_c$—mainly because $\theta_{lin}$ shows where $N$ vanishes, while $\theta_c$ the angle at which $N$ is sufficiently large to cancel out the dispersion effects.

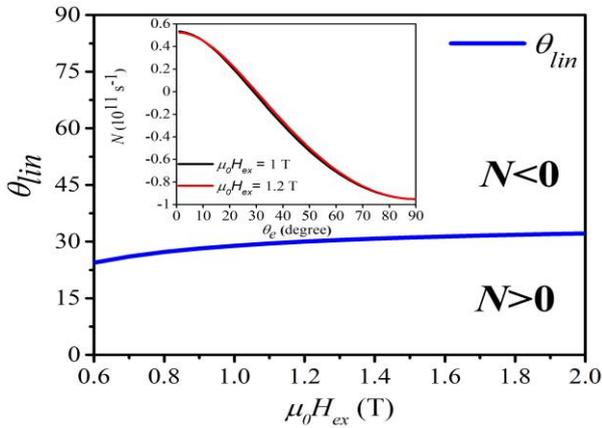

Figure 5. Field-dependent $\theta_{lin}$, where the nonlinear frequency shift parameter $N$ changes sign; Inset shows $N$ for $\mu_0 H_{ex}$ = 1 T and 1.2 T.

## IV. CONCLUSION

In conclusion, we have presented evidence of droplet nucleation in oblique fields in orthogonal STNOs. Investigation of presence of the droplet in oblique field was carried out via MR and frequency measurements, micromagnetic simulation and calculation of spin wave nonlinearity parameters. Our findings show that droplet nucleation is not limited to perpendicular fields. Indeed, the angular dependence of spin-wave dynamics in these systems shows two fundamentally different modes. While droplets nucleate in nearly perpendicular fields, propagating spin waves tend to be the dominant mode of the system in the presence of in-plane fields. Thus, a transition between these two modes takes place at an angle that is found to be distinguishable through a phase diagram between solitonic and propagating waves. The analytical results of nonlinear frequency shift parameter $N$ that qualitatively explain simulation and experimental results, shed light on the physical mechanism behind this mode conversion and point to the possible presence of the droplet in oblique fields. In addition, this shows that the nonlinearity response in PMA-STNOs has the opposite fashion to what was determined in IMA-STNOs. Our results give better insight to understanding nanoscale solitons and spin-wave dynamics in STNOs.

## ACKNOWLEDGMENTS

Majid Mohseni acknowledges support from the Iran Science Elites Federation (ISEF), the Iran National Science Foundation (INSF), the Iran Nanotechnology Initiative Council (INIC) and the Iran National Elites Foundation (INEF).

† M.M. and M.H. contributed equally to this work.

* Corresponding author email address:
m-mohseni@sbu.ac.ir, majidmohseni@gmail.com

§ Current address: Fachbereich Physik and Landesforschungszentrum OPTIMAS, Technische Universität Kaiserslautern, 67663 Kaiserslautern, Germany.

## APPENDIX A: MAGNETIC BOUNDARY CONDITION FOR PERPENDICULAR MAGNETIC ANISOTROPY SYSTEM

Total magnetic energy for a thin film ferromagnet is given by

$$W = \int w d^3 r = -\int \vec{M} \cdot \left( \vec{H}_{ex} + \frac{1}{2}\vec{H}_{dip} + \frac{1}{2}\vec{H}_k + \frac{1}{2}\vec{H}_{exch} \right) d^3 r \quad \text{(A1)}$$

Where $w$, $\vec{H}_{ex}$, $\vec{H}_{dip}$, $\vec{H}_k$ and $\vec{H}_{exch}$ are magnetic energy density, external field, dipolar (demagnetization) field, anisotropy field and exchange field, respectively, and are of the form

$$\begin{aligned}
\vec{H}_{dip} &= -\left(\vec{M} \cdot \hat{e}_z\right)\hat{e}_z \\
\vec{H}_k &= \frac{H_k}{M_{s,f}}\left(\vec{M} \cdot \hat{e}_z\right)\hat{e}_z \\
\vec{H}_{exch} &= -l_{ex}^2 \vec{\nabla}^2 \vec{M}
\end{aligned} \quad \text{(A2)}$$

The internal effective field in the ferromagnetic layer can be determined as

$$\vec{H}_{eff} = -\frac{\delta W}{\delta \vec{M}} \quad (A3)$$

In the case of static and uniform magnetization the effective field become

$$\vec{H}_{eff} = H\hat{e}_\zeta = H_{ex}\hat{e}_{ex} + (H_k - M_s)(\hat{e}_\zeta \cdot \hat{e}_z)\hat{e}_z \quad (A4)$$

where $\hat{e}_\zeta$ indicates the equilibrium magnetization direction. As indicated in Fig. 1(a) in the manuscript, the equilibrium magnetization direction and external field direction are determined by angles $\theta_f$ and $\theta_{ex}$, respectively. Azimuthal angle for magnetization direction is the same as the azimuthal angle for external field, $\varphi$, due to uniaxial perpendicular magnetic anisotropy of the free layer. Therefore, equating different components of the Eq. A4, we obtain two equation for two unknowns H and $\theta_f$ as

$$H\cos\theta_f = H_{ex}\cos\theta_{ex}$$
$$H\sin\theta_f = H_{ex}\sin\theta_{ex} + (H_k - M_s)\sin\theta_f \quad (A5)$$

By solving Eq. A5 simultaneously, one can obtain internal static field strength and direction.

## APPENDIX B: EQUATION OF MOTION FOR DIMENSIONLESS COMPLEX SPIN WAVE AMPLITUDE

Current induced magnetization dynamics in a nano-contact STO can be described by Landau-Lifshitz-Gilbert (LLG) equation with an additional Slonczewski spin transfer torque term as

$$\frac{\partial \vec{M}}{\partial t} = -\gamma\left[\vec{M}\times\vec{H}_{eff}\right] + \vec{T}_d + \vec{T}_s \quad (B1)$$

Where the Gilbert damping torque is

$$\vec{T}_d = \alpha \vec{M}\times\frac{\partial \vec{M}}{\partial t} \quad (B2)$$

and the spin transfer torque term is

$$\vec{T}_s = \frac{\hbar\varepsilon J}{M_{s,f}^2 e\mu_0 t_f}\frac{\lambda^2}{(\lambda^2+1) + (\lambda^2-1)\hat{e}_\zeta\cdot\hat{e}_p}\left(\vec{M}\times(\vec{M}\times\hat{e}_p)\right) \quad (B3)$$

All parameters in Eq. B3 are introduced in the main text except $\hat{e}_p$, which represents the polarization of the applied current through the nano-contact. One can find the polarization direction by solving Eq. A5 for the polarizer (fixed) layer with $H_k = 0$ as

$$H_p\cos\theta_p = H_{ex}\cos\theta_{ex}$$
$$H_p\sin\theta_p = H_{ex}\sin\theta_{ex} - M_{s,p}\sin\theta_p \quad (B4)$$

It is convenient to introduce the magnetization vector of the free layer in coordinate system connected with the equilibrium direction, $\hat{e}_\zeta$. In addition to $\hat{e}_\zeta$, we introduce two unit vectors $\hat{e}_\xi$ and $\hat{e}_\eta$ to form a orthogonal right-handed coordinate system as

$$\hat{e}_\zeta = \cos\theta_f\cos\varphi\hat{x} + \cos\theta_f\sin\varphi\hat{y} + \sin\theta_f\hat{z}$$
$$\hat{e}_\xi = \sin\theta_f\cos\varphi\hat{x} + \sin\theta_f\sin\varphi\hat{y} - \cos\theta_f\hat{z} \quad (B5)$$
$$\hat{e}_\eta = -\sin\varphi\hat{x} + \cos\varphi\hat{y}$$

Then, the magnetization vector of the free layer can be written as

$$\vec{M} = M_\zeta\hat{e}_\zeta + m_\xi\hat{e}_\xi + m_\eta\hat{e}_\eta$$
$$M_{s,f}^2 = M_\zeta^2 + m_\xi^2 + m_\eta^2 \quad (B6)$$

Eq. B6 implies that the magnetization has only two independent components, so we can describe it by only one complex variable

$$c(\vec{r},t) = \frac{m_\xi + im_\eta}{\sqrt{2M_{s,f}(M_{s,f} + M_\zeta)}} \quad (B7)$$

$$\vec{M} = M_{s,f}\left[(1-2|c|^2)\hat{e}_\zeta + \sqrt{1-|c|^2}\left\{(\hat{e}_\xi - i\hat{e}_\eta)c + (\hat{e}_\xi + i\hat{e}_\eta)c^*\right\}\right] \quad (B8)$$

Eq. B7 is known as Holstein-Primakoff transformation and $c(\mathbf{r},t)$ is dimensionless complex spin wave amplitude. Substituting Eq. B8 into Eq. A1 and considering magnetic energy density $w$ and LLG equation one can obtain the spin wave Hamiltonian of the form

$$\mathcal{H} = \frac{\gamma w}{2M_{s,f}} = \mathcal{A}|c|^2 + \frac{1}{2}(\mathcal{B}c^2 + c.c.) + \omega_M l_{ex}^2|\vec{\nabla}c|^2 + (\mathcal{V}|c|^2 c + c.c.) + \mathcal{U}_1|c|^4 + (\mathcal{U}_2|c|^2 c^2 + c.c.) \quad (B9)$$

where c.c. represents complex conjugate and

$$\mathcal{A} = \omega_H - \frac{1}{2}(\omega_k - \omega_M)\cos^2\theta_f \quad (B10)$$

$$\mathcal{B} = -\frac{1}{2}(\omega_k - \omega_M)\cos^2\theta_f \quad (B11)$$

$$\mathcal{V} = -(\omega_k - \omega_M)\sin\theta_f\cos\theta_f \quad (B12)$$

$$\mathcal{U}_1 = (\omega_k - \omega_M)\left(\frac{3}{2}\cos^2\theta_f - 1\right) \tag{B13}$$

$$\mathcal{U}_2 = \frac{1}{4}(\omega_k - \omega_M)\cos^2\theta_f \tag{B14}$$

$$\omega_H = \gamma H$$
$$\omega_M = \gamma M_{s,f} \tag{B15}$$
$$\omega_k = \gamma H_k$$

The quadratic part of the Hamiltonian in Eq. B9 i.e. terms including $\mathcal{A}$ and $\mathcal{B}$ can be diagonalized by a well-known $u - v$ linear canonical transformation

$$a = ub - vb^* \tag{B16}$$

Where the transformation coefficients are given by

$$u = sign(\mathcal{A})\sqrt{\frac{\mathcal{A}+\omega_0}{2\omega_0}}$$
$$v = \frac{\mathcal{B}^*}{|\mathcal{B}|}\sqrt{\frac{\mathcal{A}-\omega_0}{2\omega_0}} \tag{B17}$$

and the linear FMR frequency, $\omega_0$, is given by

$$\omega_0 = \sqrt{\mathcal{A}^2 - |\mathcal{B}|^2} = \sqrt{\omega_H\left(\omega_H - (\omega_k - \omega_M)\cos^2\theta_f\right)} \tag{B18}$$

which is used in Fig. 2 in the main text. The Hamiltonian $\mathcal{H}$ in terms of new variables $b$ and $b^*$ is

$$\mathcal{H}_b = \omega_0|b|^2 + D\left|\vec{\nabla}b\right|^2 + \left(\mathcal{W}_1|b|^2 b + c.c.\right) + \left(\mathcal{W}_2 b^3 + c.c.\right) + \mathcal{T}|b|^4 \tag{B19}$$

with

$$D = \left(u^2 + |v|^2\right)\omega_M l_{ex}^2 = \frac{\mathcal{A}}{\omega_0}\omega_M l_{ex}^2 \tag{B20}$$

$$\mathcal{W}_1 = \frac{3}{2}\left(u^2 + |v|^2\right)(u\mathcal{V} - v^*\mathcal{V}^*) - \frac{1}{2}(u\mathcal{V} + v^*\mathcal{V}^*) \tag{B21}$$

$$\mathcal{W}_2 = -uv^*(u\mathcal{V} - v^*\mathcal{V}^*) \tag{B22}$$

$$\mathcal{T} = \frac{1}{2}\left[3\left(u^2 + |v|^2\right)^2 - 1\right]\mathcal{U}_1 - 3u\left(u^2 + |v|^2\right)\left(v\mathcal{U}_2 + v^*\mathcal{U}_2^*\right) \tag{B23}$$

We can eliminate the nonresonant three-wave processes by introducing the nonlinear transformation with new variables $a$ and $a^*$ as

$$b \approx \sqrt{\frac{\omega_0}{\mathcal{A}}}a + \frac{\omega_0}{\mathcal{A}}\frac{\mathcal{W}_1 a^2 - 2\mathcal{W}_1^*|a|^2 - \mathcal{W}_2^* a^{*2}}{\omega_0} + \mathcal{O}(a^3) \tag{B24}$$

The Hamiltonian $\mathcal{H}_a = \mathcal{A}\mathcal{H}_b/\omega_0$ in terms of variables $a$ and $a^*$ is given by

$$\mathcal{H}_a = \omega_0|a|^2 + D\left|\vec{\nabla}a\right|^2 + \frac{N}{2}|a|^4 \tag{B25}$$

where $D$ is known as dispersion coefficient. Nonlinear frequency shift coefficient, $N$, is

$$N = \frac{2\omega_0}{\mathcal{A}}\left(\mathcal{T} - 3\frac{|\mathcal{W}_1|^2 + |\mathcal{W}_2|^2}{\omega_0}\right) \tag{B26}$$

which is used to provide Fig. 5 in the main text. Applying the mentioned transformations, one can obtain the LLG-S equation (Eq. B1) in terms of complex spin wave variables $a$ and $a^*$ as

$$\frac{\partial a}{\partial t} = i\frac{\delta \mathcal{H}_a}{\delta a} + F_d + F_J \tag{B27}$$

where $F_d$ and $F_J$ represent forces due to damping and spin torque. Therefore, the equation of motion for the dimensionless complex spin wave amplitude, $a(\mathbf{r}, t)$, is

$$i\frac{\partial a}{\partial t} = -\left(\omega_0 - D\nabla^2 + N|a|^2\right)a + i\left(\Gamma_J - \Gamma_0 + \kappa\nabla^2\right)a. \tag{B28}$$

All the parameters appeared in the Eq. B28 are introduced in the main text.